\documentclass[11pt]{article}

\usepackage{amsmath, amsfonts}

\usepackage{color}
\usepackage{algorithm, algpseudocode, algorithmicx}
\usepackage{verbatim}
\usepackage{setspace}

\usepackage[dvips]{graphicx}
\graphicspath{{./eps/}}
\DeclareGraphicsExtensions{.eps}
\usepackage{subfig}

\title{A Bayesian Approach to the Partitioning of Workflows}

\author{
	Freddy C. Chua and Bernardo A. Huberman\\
	Mechanisms and Design Lab, Hewlett Packard Labs
}

\begin{document}

\maketitle

\begin{abstract}
When partitioning workflows in realistic scenarios, the knowledge of the processing units is often vague or unknown. A naive approach to addressing this issue is to perform many controlled experiments for different workloads,  each consisting of multiple number of trials in order to estimate the mean and variance of the specific workload. Since this controlled experimental approach can be quite costly in terms of time and resources, we propose a variant of the Gibbs Sampling algorithm that uses a sequence of Bayesian inference updates to estimate the processing characteristics of the processing units. Using the inferred characteristics of the processing units, we are able to determine the best way to split a workflow for processing it in parallel with the lowest expected completion time and least variance. 
\end{abstract}

\section{Introduction}

Many large and time consuming tasks can be broken down into  independent components, as for example, $i$ and $j$, with proportions of $f$ and $1 - f$ for processing in parallel \cite{Huberman2015}. The task is considered complete when both independent components complete, with the processing time taken as the maximum of the two components. Given that each component operates on distinct processing unit with different configurations and capabilities, each component has a completion time that follows a different statistical distribution. If we let $\Theta_i$ represent the parameters of $i$'s completion time $t_i$, and $\Theta_j$ to represent the parameters of $j$'s completion time $t_j$, 
\begin{enumerate}
	\item The probability that the task has a completion time $t$ before $\epsilon$ is given by,
		\begin{align*}
			P(t \leq \epsilon | f, \Theta) &= P(t_i \leq \epsilon | f, \Theta_i) P(t_j \leq \epsilon | f, \Theta_j) \\
			\Theta &= \{ \Theta_i, \Theta_j \}
		\end{align*}
	\item The expected completion time of the task $t$ is given by,
		\[ E(t | f, \Theta) = \int_0^{\infty} 1 - P(t \leq \epsilon | f, \Theta) ~ d\epsilon \]
	\item While the variance of completion time $t$ is given by,
		\[ Var(t | f, \Theta) = \left\{ 2 \int_0^{\infty} \epsilon \Big[ 1 - P(t \leq \epsilon | f, \Theta) \Big] ~ d\epsilon \right\} - \Big[ E(t | f, \Theta) \Big]^2 \]
\end{enumerate}

For brevity, we shall denote the expected completion time $E(t|f,\Theta)$ as $\mu(f)$ and the variance of completion time $Var(t|f,\Theta)$ as $\sigma^2(f)$.

\begin{figure}[htb]
	\centering
	\includegraphics[width=4.0in]{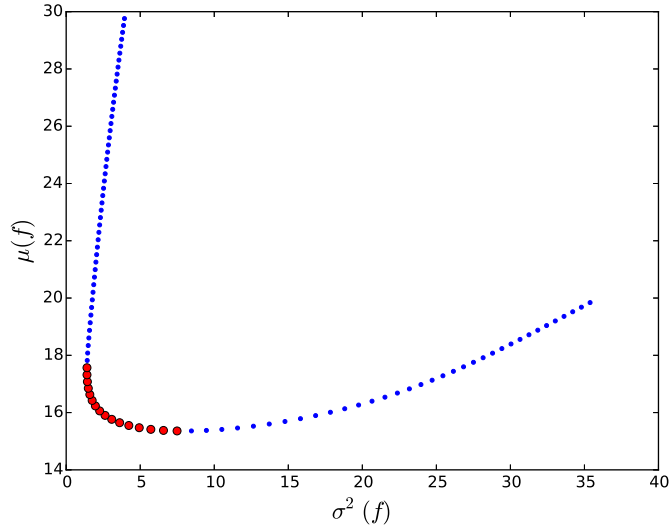}
	\caption{$\mu(f)$ and $\sigma^2(f)$ for each value of $f$}
	\label{fig:mu_vs_sigma2}
\end{figure}

\begin{figure}[htb]
	\centering
	\subfloat[$\mu(f)$ with respect to $f$]
	{
		\includegraphics[width=2.3in]{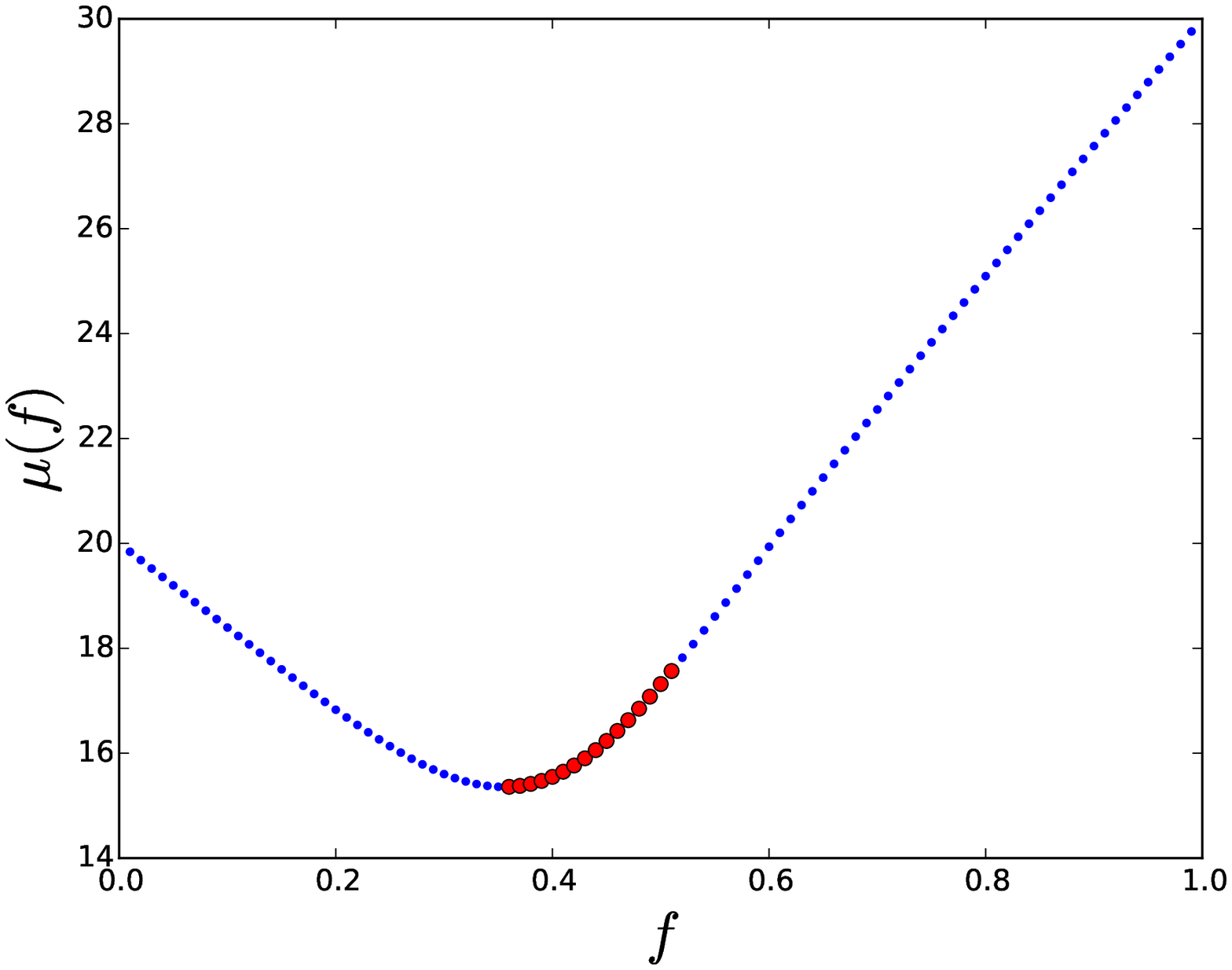}
		\label{fig:mu_vs_f}
	}
	\subfloat[$\sigma^2(f)$ with respect to $f$]
	{
		\includegraphics[width=2.3in]{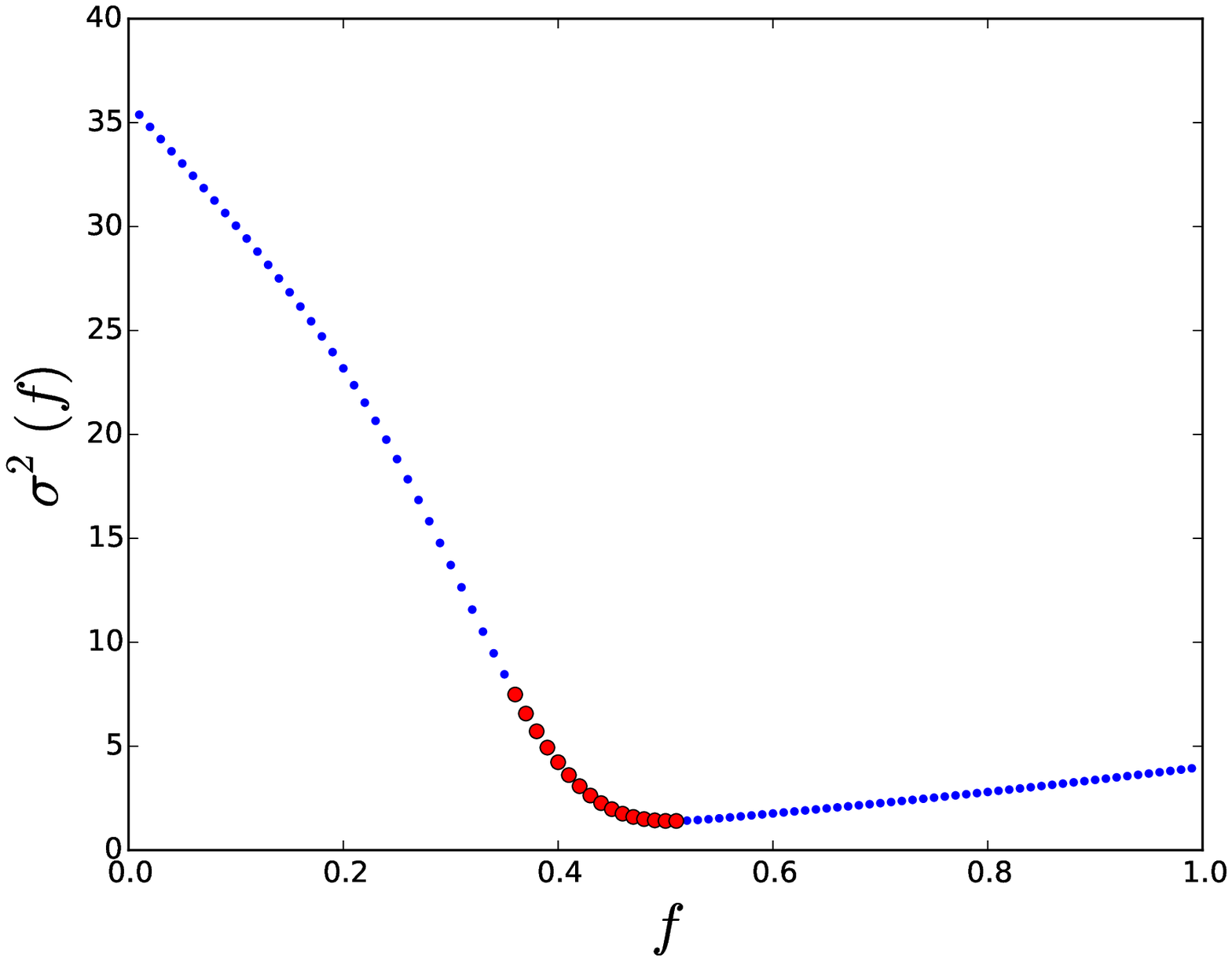}
		\label{fig:sigma2_vs_f}
	}
	\caption{$\mu(f)$ and $\sigma^2(f)$ as a function of $f$}
	\label{fig:portfolio}
\end{figure}

For the purpose of illustration on how $\mu(f)$ and $\sigma^2(f)$ vary as a function of $f$, we shall assume that the completion times of each processing unit is Gaussian in nature, with known values of the parameter $\Theta$, which governs the processing capabilities of the two processing units. By using the hypothetical values $\mu_i = 30$, $\sigma_i = 2$, $\mu_j = 20$, $\sigma_j = 6$, we obtain the numerical results as shown in Figures \ref{fig:mu_vs_sigma2}, \ref{fig:mu_vs_f} and \ref{fig:sigma2_vs_f}.

In Figure \ref{fig:mu_vs_sigma2}, each point gives the respective value of $\mu(f)$ and $\sigma^2(f)$ for a specific $f$. The curve formed from these points is parabolic which indicates that for some values of $\mu(f)$, there are two possible choices of $\sigma^2(f)$. The converse is true as well, i.e. some values of $\sigma^2(f)$ has two choices of $\mu(f)$. If our assumptions on the statistical distribution of completion times for the two parallel machines hold, then the theoretical results derived in Figure \ref{fig:mu_vs_sigma22} allows us to decide the appropriate values of $f$ which,
\begin{enumerate}
	\item Minimizes the expected time $\mu(f)$ for a desired amount of uncertainty $\sigma^2(f)$.
	\item Minimizes the amount of uncertainty $\sigma^2(f)$ for a desired expected time $\mu(f)$.
\end{enumerate}
The appropriate choice of $f$ that provides the optimal values of $\mu(f)$ and $\sigma^2(f)$ is given by the efficient frontier in the lower left portion of the curve which is highlighted in \textcolor{red}{bolded red $\bullet$}. $f$ thus denote the amount of appropriate parallelism necessary to achieve a desired level of Quality-of-Service (QoS).

The scenarios for which QoS  is important are supply chain management, computer networking, parallel and distributed systems, or even military strategies that often require the achievement of a common objective orchestrated by several teams working in parallel.

\subsection*{Problem Description}

But in many realistic scenarios, the knowledge of the processing unit is often vague or unknown. A \textbf{naive} approach to addressing this issue is to perform many controlled experiments of different workloads by varying $f$, with each $f$ having multiple number of trials in order to estimate the mean and variance at each value of $f$. However, in realistic or deployed systems, such controlled experiments represent an opportunity cost and the resources used to conduct such experiments would reap more benefits by running actual workloads.

It is therefore necessary to have an algorithm which learns the (processing) system parameters quickly based on several trials of using the processing units without deliberate selection of $f$. To fulfill this requirement, we propose to use a Bayesian approach to infer the parameters $\Theta$. Using a Bayesian approach allows several benefits as follows,
\begin{enumerate}
	\item The current understanding of the systems' performance can be given as input to the algorithm using prior beliefs expressed using statistical distributions.
	\item Based on an observed batch of data, such as completion time with respect to the amount of parallelism $f$, the likelihood of the observations can be combined with the prior beliefs to obtain a posterior belief of the systems' performance.
	\item The posterior belief obtained from the previous batch of observations can become the prior belief for the next batch of observations. By chaining a sequence of prior and posterior updates, the algorithm can adjust the systems' parameters for a dynamically fast changing environment.
\end{enumerate}

\section{The Splitting Workflow Model based on the Normal distribution}

For simplification (but without loss of generality) of the learning algorithm description, we shall assume that the completion time $t_i, t_j$ for each of the processing unit $i,j$ follows that of a Normal distribution.
\begin{gather*}
	p(t_i | f, \mu_i, \sigma_i, \alpha_i, \beta_i) \sim \mathcal{N} \left( f^{\alpha_i} \mu_i, \left[ f^{\beta_i} \sigma_i \right]^2 \right) \\
	p(t_j | f, \mu_j, \sigma_j, \alpha_j, \beta_j) \sim \mathcal{N} \left( [1-f]^{\alpha_j} \mu_j, \left[ (1-f)^{\beta_j} \sigma_j \right]^2 \right)
\end{gather*}
where $\alpha$ and $\beta$ are scaling exponents that affects the completion time for varying size of the workload. In efficient and ideal parallel systems, $\alpha$ and $\beta$ would have values of $1.0$. But due to coordination costs and communication overheads in parallel processing, the values of $\alpha$ and $\beta$ are unlikely to have an exact value of $1.0$. Since the $\alpha$ and $\beta$ have an inter-dependency with the values of $\mu$ and $\sigma$, that implies estimating the parameters of the model cannot be easily reduce to estimating the parameters of a Normal distribution.

Let's simplify the notation so that,
\begin{align*}
	f_i &= f \\
	f_j &= 1 - f_i
\end{align*}
Then we can see that the analysis for $i$ and $j$ is identical,
\begin{align*}
	p(t_i | f_i, \mu_i, \sigma_i) &\sim \mathcal{N} \left( f_i^{\alpha_i} \mu_i, f_i^{2\beta_i} \sigma_i^2 \right) \\
	p(t_j | f_j, \mu_j, \sigma_j) &\sim \mathcal{N} \left( f_j^{\alpha_j} \mu_j, f_j^{2\beta_j} \sigma_j^2 \right)
\end{align*}

The purpose of simplifying for $i$ and $j$ is to show that if we can derive the bayesian updates of $i$, then we can similarly apply the same equations to $j$. With that, we can reduce the clutter in the equations by dropping the subscripts $i$ and $j$ so that we only have to work on the following, 
\begin{align}
	\label{eqn:model_t}
	p(t|f, \mu, \sigma, \alpha, \beta) \sim \mathcal{N} \left( f^{\alpha} \mu, f^{2\beta} \sigma^2 \right)
\end{align} 
As stated in Equation \ref{eqn:model_t}, the completion time $t$ can be predicted conditioned on the assumption that $\mu,\sigma, \alpha, \beta$ are known. The original motivation of our discussion does not assume knowledge of these values. 

In the next few sections, we will derive the Bayesian inference equations that allow us to obtain estimations for the values of $\mu, \sigma, \alpha, \beta$.

\section{Bayesian Inference for $\mu$ and $\sigma$}

Since these values are unknowns, we can assume that they are drawn from some statistical distributions. For notational convenience, let's replace the variance $\sigma^2$ with the precision $\lambda$ using the following relationship,
\[ \lambda = \frac{1}{\sigma^2} \]
An appropriate choice of prior distribution for $\mu$ is the following Normal distribution,
\begin{align*}
\mu | \mu_0, \kappa_0, \lambda &\sim \mathcal{N} \left( \mu_0, ( \kappa_0 \lambda )^{-1} \right) \\
p(\mu | \mu_0, \kappa_0, \lambda) &\propto \lambda^{\frac{1}{2}} \exp\left( -\frac{\kappa_0 \lambda}{2} (\mu-\mu_0)^2 \right)	
\end{align*}
While the prior distribution of $\lambda$ is the Gamma distribution,
\begin{align*}
	\lambda | \nu_0, \psi_0 &\sim \text{Gamma}(\nu_0, \text{rate}=\psi_0) \\
	p(\lambda | \nu_0, \psi_0) &\propto \lambda^{\nu_0 - 1} \exp \left( -\psi_0 \lambda \right)
\end{align*}
where $\mu_0, \kappa_0, \nu_0$ and $\psi_0$ are parameters for the prior distributions of $\mu$ and $\lambda$, which are constants that can be set based on subjective prior knowledge.

Then expressing the pdf as a multiplication of the two distribution, 
\begin{align}
	p(\mu, \lambda | \mu_0, \kappa_0, \nu_0, \psi_0) &\propto \lambda^{\frac{1}{2}} \exp \left( -\frac{\kappa_0 \lambda}{2} (\mu - \mu_0)^2 - \psi_0 \lambda \right) \lambda^{\nu_0 - 1} \nonumber \\
		\label{eqn:mu_sigma_prior}
		&\propto \lambda^{\nu_0 - \frac{1}{2}} \exp \left( -\frac{\lambda}{2} \left[ \kappa_0 (\mu - \mu_0)^2 + 2 \psi_0 \right] \right)
\end{align}

The next step is to merge the prior distribution with the likelihood of some observed data to obtain the posterior distribution. In statistical notation, we would like to obtain the posterior distribution conditioned on the observations of some completion time $T = \{t_1, t_2, \ldots, t_N\}$ for a given set of workload $F = \{f_1, f_2, \ldots, f_N \}$. And assuming that the values of $\alpha$ and $\beta$ is known. i.e.
\begin{gather}
	\label{eqn:mu_sigma_posterior}
	p(\mu, \lambda | T, F, \alpha, \beta, \mu_0, \kappa_0, \nu_0, \psi_0) \propto p(T | F, \mu, \lambda, \alpha, \beta) p(\mu, \lambda | \mu_0, \kappa_0, \nu_0, \psi_0)
\end{gather}
The likelihood is then given by,
\begin{align}
	p(T | F, \mu, \lambda, \alpha, \beta) = &\prod_n p(t_n | f_n, \mu, \lambda, \alpha, \beta) \nonumber \\
	p(t_n | f_n, \mu, \lambda, \alpha, \beta) &\propto \frac{\sqrt{\lambda}}{f_n^\beta} \exp \left( -\frac{\lambda}{2} \left[ \frac{t - f_n^\alpha \mu}{f_n^\beta} \right]^2 \right) \nonumber \\
	p(T | F, \mu, \lambda, \alpha, \beta) &\propto \prod_n \frac{\sqrt{\lambda}}{f_n^\beta} \exp \left(-\frac{\lambda}{2} \left[ \frac{t - f_n^\alpha \mu}{f_n^\beta} \right]^2 \right) \nonumber \\
	\label{eqn:mu_sigma_likelihood_0}
	&\propto \frac{\lambda^{\frac{N}{2}}}{\prod_n f_n^\beta} \exp \left( - \frac{\lambda}{2} \sum_n \left[ \frac{t - f_n^\alpha \mu}{f_n^\beta} \right]^2 \right) \\
	\label{eqn:mu_sigma_likelihood}
	&\propto \lambda^{\frac{N}{2}} \exp \left( - \frac{\lambda}{2} \sum_n \left[ \frac{t - f_n^\alpha \mu}{f_n^\beta} \right]^2 \right)
\end{align}

Substitute Equations \ref{eqn:mu_sigma_prior} and \ref{eqn:mu_sigma_likelihood} into \ref{eqn:mu_sigma_posterior}. Then through some algebraic manipulations (expansion, completing the square, factorization and simplification), we can obtain the posterior distribution given by,
\begin{align*}
	&p(\mu, \lambda|T,F,\alpha,\beta,\mu_0,\kappa_0,\nu_0,\psi_0) \\
	&\propto \lambda^{\nu_N - \frac{1}{2}} \exp \left( -\frac{\lambda}{2} \left[ \kappa_N (\mu - \mu_N)^2 + 2 \psi_N \right] \right)
\end{align*}
With $\mu_N, \kappa_N, \nu_N$ and $\psi_N$ given by,
\begin{align}
	\label{eqn:mu_N}
	\mu_N &= \frac{\mu_0 \kappa_0 + \sum_n f_n^{\alpha - 2\beta} t_n}{\kappa_0 + \sum_n f_n^{2\alpha - 2\beta}} \\
	\label{eqn:kappa_N}
	\kappa_N &= \kappa_0 + \sum_n f_n^{2\alpha - 2\beta} \\
	\label{eqn:nu_N}
	\nu_N &= \nu_0 + \frac{N}{2} \\
	\label{eqn:psi_N}
	\psi_N &= \psi_0 + \frac{1}{2} \left[ -\mu_N^2 \kappa_N + \mu_0^2 \kappa_0 + \sum_n \left( \frac{t_n}{f_n^{\beta}} \right)^2 \right]
\end{align}

\section{Bayesian Inference for $\alpha$ and $\beta$}

$\alpha$ and $\beta$ represents the scalability of the processing unit when given different workloads governed by $f$. A perfect system would have $\alpha = 1.0$ and $\beta=1.0$ indicating that the expected completion time and variance scales linearly with the size of the workload. $\alpha > 1.0$ and $\beta > 1.0$ represents an impossible scenario since this suggests that the system takes less time and have less uncertainty when given more workload. Since $\alpha$ and $\beta$ could only take values between $0$ and $1$, it would be appropriate to use the Beta distribution as the prior of $\alpha$ and $\beta$.
\begin{align*}
	p(\alpha | \theta_0, \phi_0) &\propto \alpha^{\theta_0 - 1} (1 - \alpha)^{\phi_0 - 1} \\
	p(\beta  | \delta_0, \eta_0) &\propto \beta^{\delta_0 - 1} (1-\beta)^{\eta_0 - 1}
\end{align*}

Using the likelihood given by Equation \ref{eqn:mu_sigma_likelihood}, the posterior distribution of $\alpha$ conditioned on a set of observations $T$ for a given set of $F$ is,
\begin{align}
	&p(\alpha | T, F, \mu, \lambda, \theta_0, \phi_0, \beta) \propto p(T | F, \mu, \lambda, \beta, \alpha) ~ p(\alpha | \theta_0, \phi_0) \nonumber \\
	\label{eqn:alpha_posterior}
	&\propto \lambda^{\frac{N}{2}} \exp \left( - \frac{\lambda}{2} \sum_n \left[ \frac{t - f_n^\alpha \mu}{f_n^\beta} \right]^2 \right) \alpha^{\theta_0 - 1} (1 - \alpha)^{\phi_0 - 1}
\end{align}
For the posterior distribution of $\beta$, we would have to use the likelihood given by Equation \ref{eqn:mu_sigma_likelihood_0} which gives us the following,
\begin{align}
	&p(\beta | T, F, \mu, \lambda, \delta_0, \eta_0, \alpha) \propto p(T | F, \mu, \lambda, \alpha, \beta) ~ p(\beta | \delta_0, \eta_0) \nonumber \\
	\label{eqn:beta_posterior}
	&\propto \frac{\lambda^{\frac{N}{2}}}{\prod_n f_n^\beta} \exp \left( - \frac{\lambda}{2} \sum_n \left[ \frac{t - f_n^\alpha \mu}{f_n^\beta} \right]^2 \right) \beta^{\delta_0 - 1} (1-\beta)^{\eta_0 - 1}
\end{align}

\begin{figure}[htb]
	\centering
	\subfloat[True posterior distribution of $\alpha$ as given by Equation \ref{eqn:alpha_posterior}]
	{
		\includegraphics[width=2.3in]{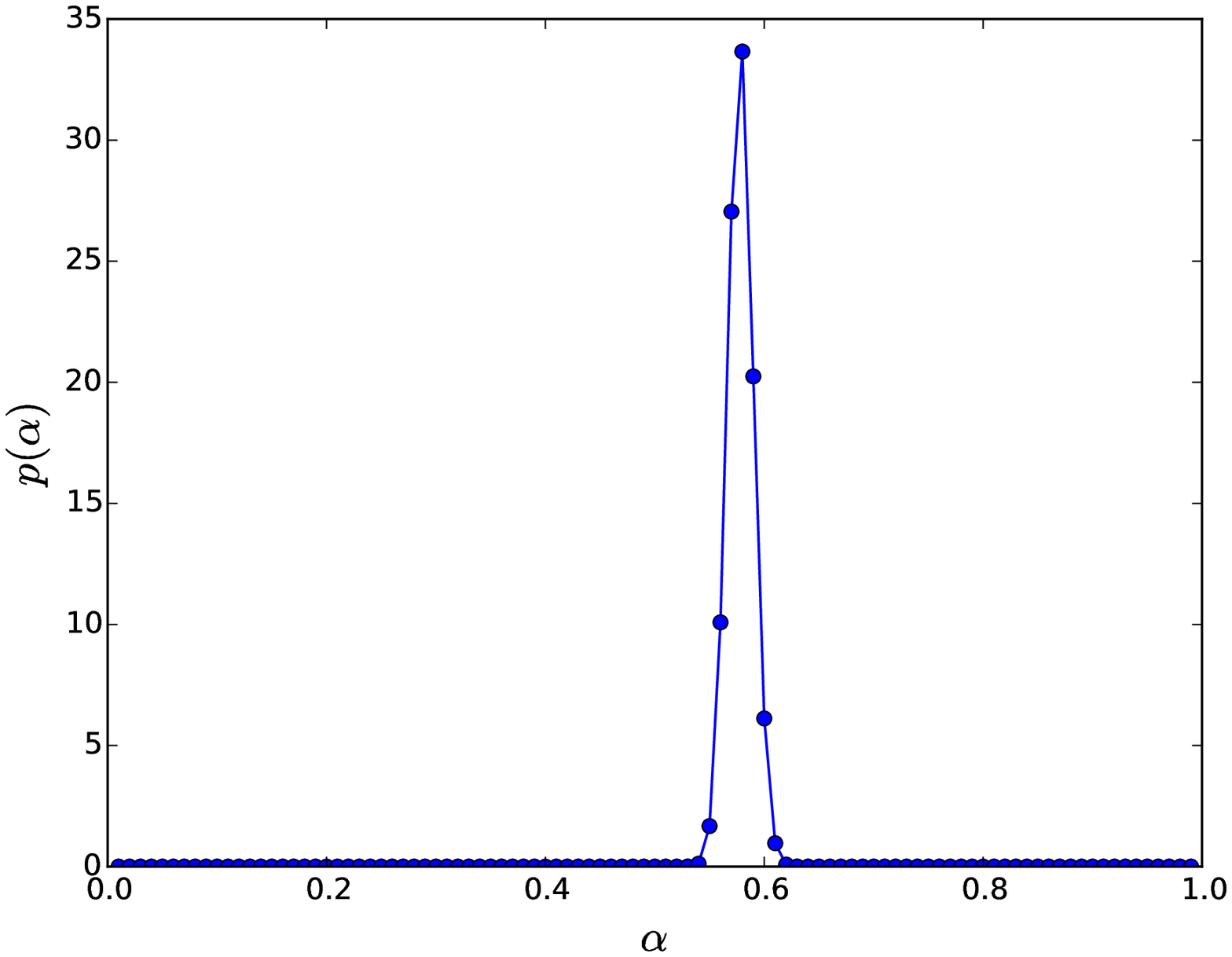}
		\label{fig:true_posterior_alpha}
	}
	\subfloat[Approximate posterior distribution of $\alpha$ using the method of moments]
	{
		\includegraphics[width=2.3in]{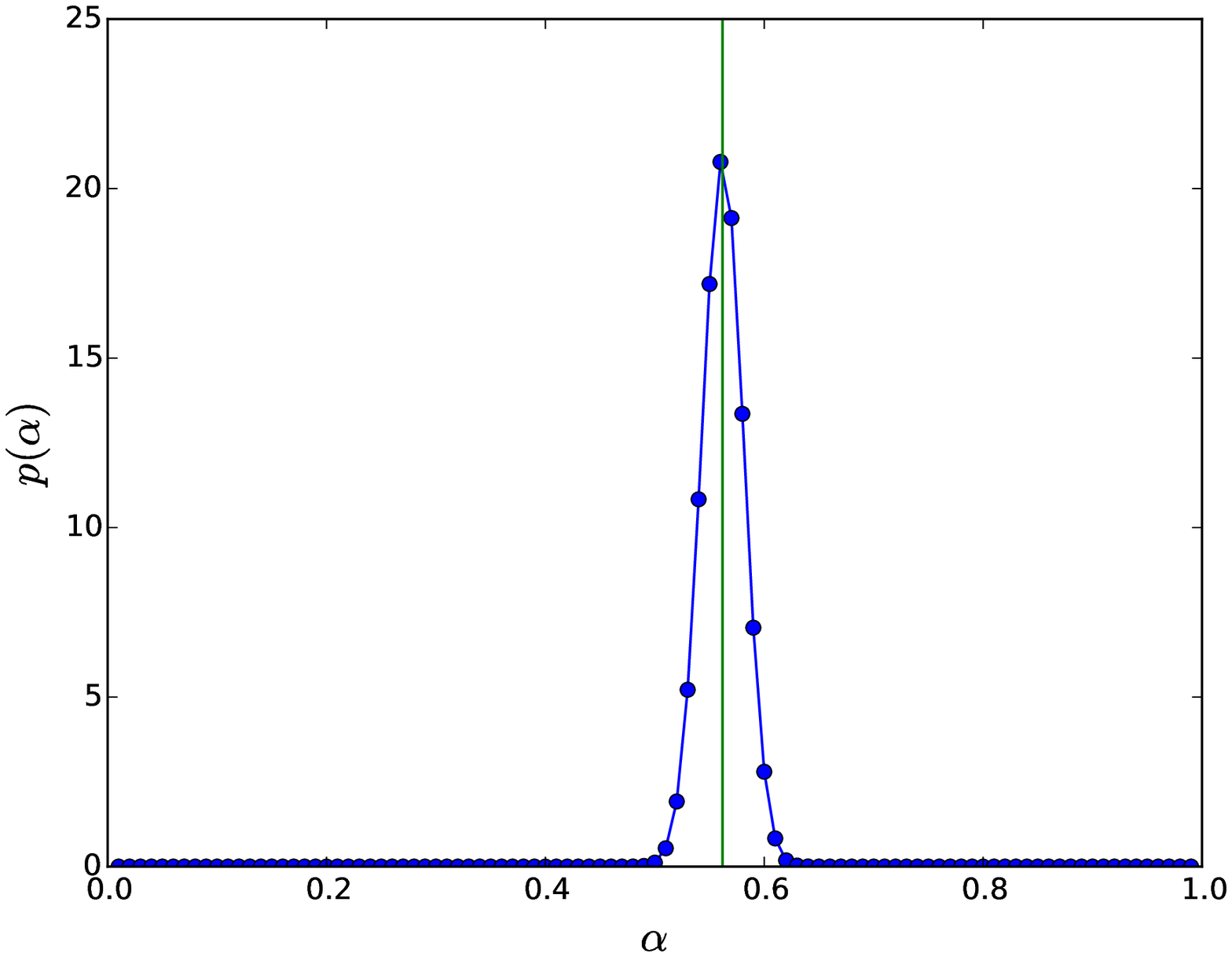}
		\label{fig:approx_posterior_alpha}
	}
	\caption{Comparison between the true and approximate posterior distribution of $\alpha$}
	\label{fig:posterior_alpha}
\end{figure}

\begin{figure}[htb]
	\centering
	\subfloat[True posterior distribution of $\beta$ as given by Equation \ref{eqn:beta_posterior}]
	{
		\includegraphics[width=2.3in]{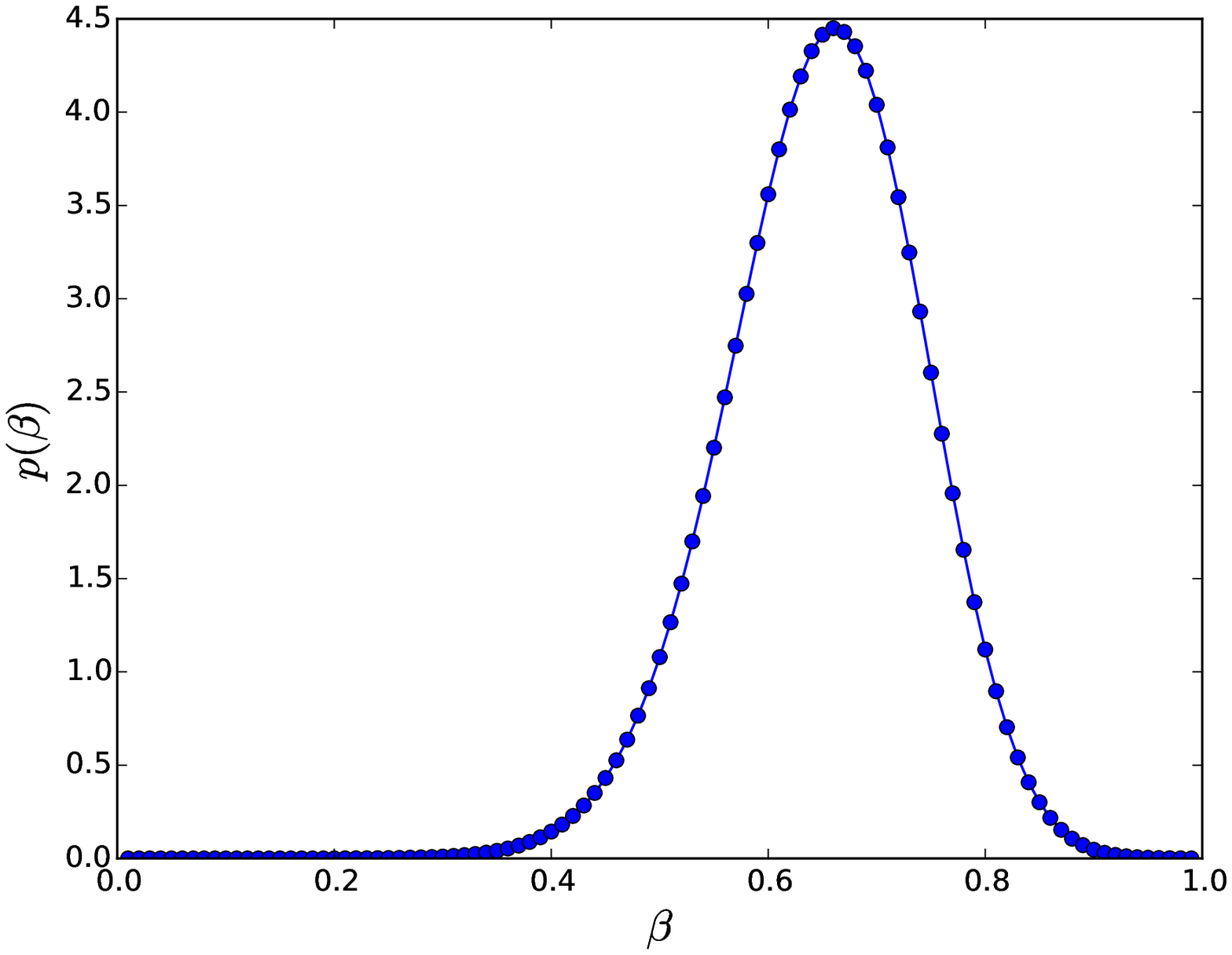}
		\label{fig:true_posterior_beta}
	}
	\subfloat[Approximate posterior distribution of $\beta$ using the method of moments]
	{
		\includegraphics[width=2.3in]{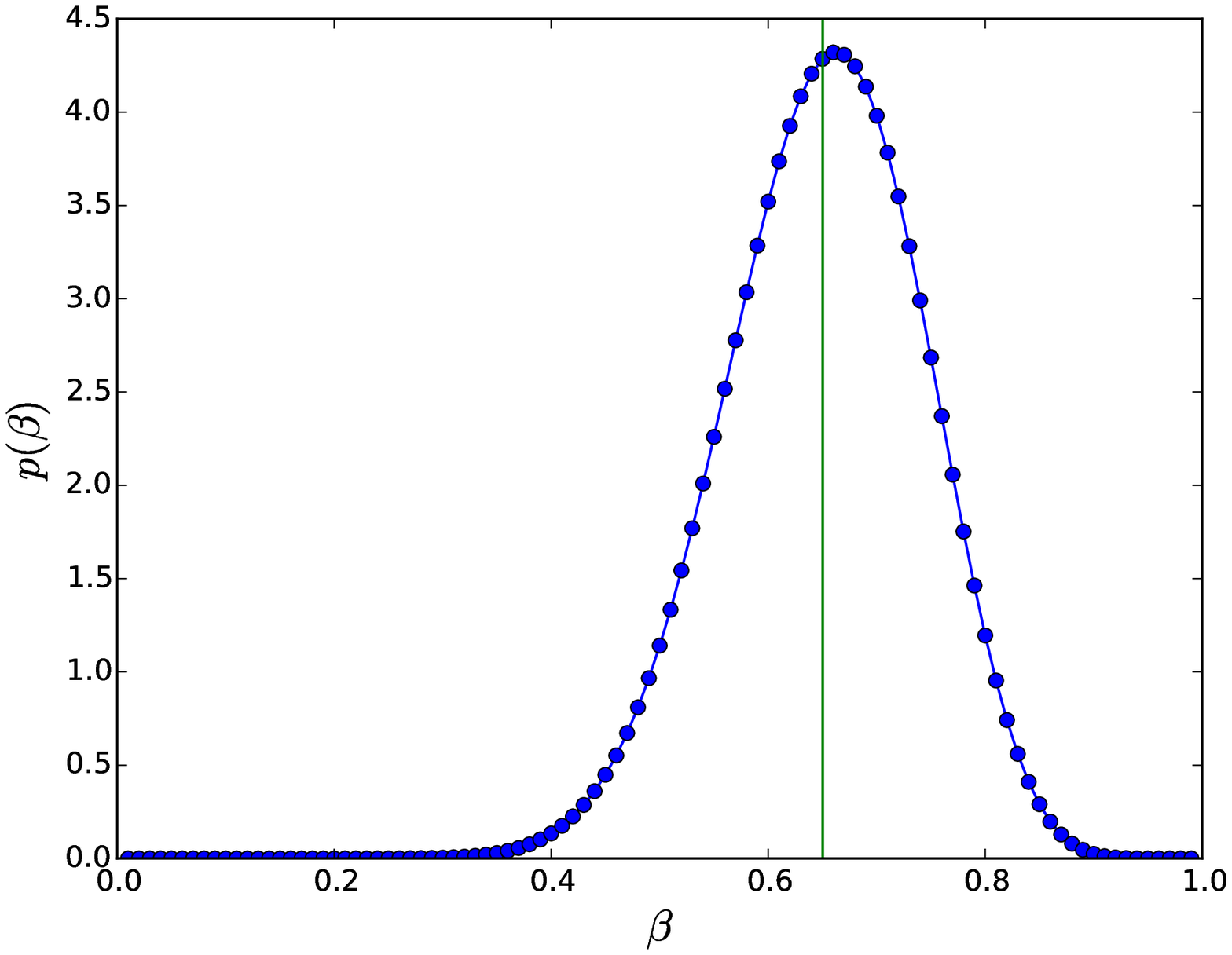}
		\label{fig:approx_posterior_beta}
	}
	\caption{Comparison between the true and approximate posterior distribution of $\beta$}
	\label{fig:posterior_beta}
\end{figure}

Unfortunately, there is no algebraic solution to manipulate the posterior distribution given by Equations \ref{eqn:alpha_posterior} and \ref{eqn:beta_posterior} into Beta distributions. In fact, there is no analytical proof that the posterior distributions remains as Beta distributions.

We could continue to assume that the posterior distribution can be approximated by a Beta distribution with parameters $\theta_N, \phi_N$ for $\alpha$ and $\delta_N, \eta_N$ for $\beta$. Using the method of moments, 
\begin{align}
	\label{eqn:theta_N}
	\theta_N &= E(\alpha) \left[ \frac{E(\alpha)[1 - E(\alpha)]}{Var(\alpha)} - 1 \right] \\
	\label{eqn:phi_N}
	\phi_N &= [1 - E(\alpha)] \left[ \frac{E(\alpha)[1 - E(\alpha)]}{Var(\alpha)} - 1 \right] \\
	\label{eqn:delta_N}
	\delta_N &= E(\beta) \left[ \frac{E(\beta)[1 - E(\beta)]}{Var(\beta)} - 1 \right] \\
	\label{eqn:eta_N}
	\eta_N &= [1 - E(\beta)] \left[ \frac{E(\beta)[1 - E(\beta)]}{Var(\beta)} - 1 \right] 
\end{align}
Then using the standard definitions for $E(\alpha)$ and $Var(\alpha)$ to derive their specific values,
\begin{align}
	\label{eqn:expected_alpha}
	E(\alpha) &= \int_0^1 \alpha \cdot p(\alpha | T, F, \mu, \lambda, \theta_0, \phi_0) ~ d\alpha \\
	\label{eqn:expected_alpha2}
	E(\alpha^2) &= \int_0^1 \alpha^2 \cdot p(\alpha | T, F, \mu, \lambda, \theta_0, \phi_0) ~ d\alpha \\
	\label{eqn:variance_alpha}
	Var(\alpha) &= E(\alpha^2) - \left[ E(\alpha) \right]^2
\end{align}
Although unproven, it is unlikely that the integrals due to the PDF given by Equation \ref{eqn:alpha_posterior} and \ref{eqn:beta_posterior} have closed form solutions. In our solver, we employ the use of numerical integration to obtain an approximate value for the expectations and variances. Similar procedure applies for $\delta_N$ and $\eta_N$ of $\beta$.

Figures \ref{fig:true_posterior_alpha} and \ref{fig:approx_posterior_alpha} show an example of the differences between the true and approximate posterior distribution of $\alpha$. Figures \ref{fig:true_posterior_beta} and \ref{fig:approx_posterior_beta} show an example of the differences between the true and approximate posterior distribution of $\beta$. The green line in Figures \ref{fig:approx_posterior_alpha} and \ref{fig:approx_posterior_beta} shows that the mean of the distribution is also close to the mode of the distribution, which has important implications for the Gibbs Sampling algorithm which we will describe in the next section.

\section{Gibbs Sampling Algorithm}

Algorithm \ref{alg:gibbs_sampling} summarizes the use of the Bayesian inference equations for estimating the parameters of the processing system. After updating the parameters of the prior distributions, we \textbf{sample} from the distributions instead of taking their mode or mean so as to avoid getting trapped in a local maxima of the log likelihood. Due to the fact that the mean is also closed to the mode as shown in Figures \ref{fig:approx_posterior_alpha} and \ref{fig:approx_posterior_beta}, it suggest that sampling from their distributions will have the desired side effect of increasing the log likelihood of the overall system.

\begin{algorithm}[htb]
	\caption{Gibbs Sampling of $\mu, \sigma, \alpha, \beta$}
	\label{alg:gibbs_sampling}
	\begin{algorithmic}
		\Require $\{ \mu_0, \kappa_0, \nu_0, \psi_0 \}, \{ \theta_0, \phi_0 \}, \{ \delta_0, \eta_0 \}$ 
		\State Sample $\alpha$ from Beta distribution using $\theta_0$ and $\phi_0$.
		\State Sample $\beta$ from Beta distribution using $\delta_0$ and $\eta_0$.
		\While{true}
			\State $T \leftarrow$ [], $F \leftarrow$ []
			\For{$n \leftarrow 1$ to $N$}
				\State Add $t_n$ to $T$, add $f_n$ to $F$
			\EndFor
			\For{some number of iterations}
				\State $\mu_N \leftarrow$ using Equation \ref{eqn:mu_N}.
				\State $\kappa_N \leftarrow$ using Equation \ref{eqn:kappa_N}.
				\State $\nu_N \leftarrow$ using Equation \ref{eqn:nu_N}.
				\State $\psi_N \leftarrow$ using Equation \ref{eqn:psi_N}.
				
				\State Sample $\lambda$ from Gamma distribution using $\nu_N$ and $\psi_N$.
				\State Sample $\mu$ from Normal distribution using $\mu_N$ and $(\kappa_N \lambda)^{-1}$.
				\State $\theta_N \leftarrow$ using Equation \ref{eqn:theta_N}.
				\State $\phi_N \leftarrow$ using Equation \ref{eqn:phi_N}.
				\State Sample $\alpha$ from Beta distribution using $\theta_N$ and $\phi_N$.
				\State $\delta_N \leftarrow$ using Equation \ref{eqn:delta_N}.
				\State $\eta_N \leftarrow$ using Equation \ref{eqn:eta_N}.
				\State Sample $\beta$ from Beta distribution using $\delta_N$ and $\eta_N$.
			\EndFor
			\State $\mu_0 \leftarrow \mu_N$, $\kappa_0 \leftarrow \kappa_N$, $\nu_0 \leftarrow \nu_N$, $\phi_0 \leftarrow \phi_N$
			\State $\theta_0 \leftarrow \theta_N$, $\phi_0 \leftarrow \phi_N$, $\delta_0 \leftarrow \delta_N$, $\eta_0 \leftarrow \eta_N$
		\EndWhile
		\State $\sigma \leftarrow \sqrt{1/\lambda}$\\
		\Return $\mu, \sigma, \alpha, \beta$
	\end{algorithmic}
\end{algorithm}

\begin{figure}[htb]
	\centering
	\includegraphics[width=4.0in]{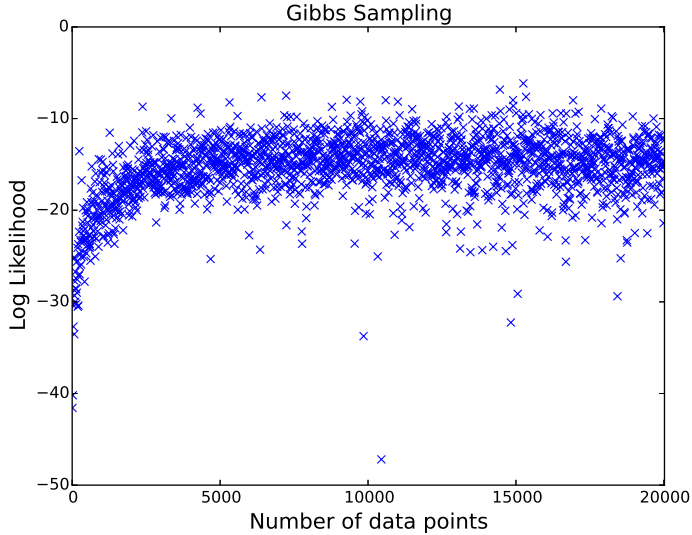}
	\caption{Results for estimating the network characteristics for a file transfer. Each point represents the logarithm likelihood of Equation \ref{eqn:mu_sigma_likelihood_0}.}
	\label{fig:digital_ocean_gibbs}
\end{figure}

Figure \ref{fig:digital_ocean_gibbs} shows the convergence of the Gibbs Sampling algorithm presented in Algorithm \ref{alg:gibbs_sampling}. The fast increase of the log likelihood (y-axis) using relatively low number of data points (x-axis) shows that the Bayesian inference equations and the Gibbs Sampling algorithm is able to estimate the system parameters.

\bibliographystyle{unsrt}
\bibliography{bayes}

\end{document}